\documentclass[a4paper,12pt]{article}

\usepackage{amsmath,amssymb,latexsym}
\newcommand{\ct}{\theta}
\newcommand{\f}{\phi}
\newcommand{\ftilde}{\tilde{\phi}}
\newcommand{\kp}{\kappa}
\newcommand{\del}{\partial}
\newcommand{\xhat}{\hat{x}}

\newcommand{\Phat}{\hat{P}}

\newcommand{\Jhat}{\hat{J}}
\begin{document}
\renewcommand{\thefootnote}{\fnsymbol{footnote}}
\begin{titlepage}
\vspace*{0mm}
\hfill
{\hfill \begin{flushright} 
YITP-07-78 
\end{flushright}
  }

\vspace*{10mm}

\begin{center}
{\LARGE {\LARGE  
Domain wall solitons and Hopf algebraic translational symmetries in 
noncommutative field theories}}
\vspace*{15mm}

 {\large Yuya Sasai}
\footnote{ e-mail: sasai@yukawa.kyoto-u.ac.jp}
 {\large and~ Naoki Sasakura}
\footnote{ e-mail: sasakura@yukawa.kyoto-u.ac.jp}

\vspace*{7mm}
{\large {\it 
Yukawa Institute for Theoretical Physics, Kyoto University, \\ 
 Kyoto 606-8502, Japan 
}} \\

\end{center}

\vspace*{1.3cm}

\begin{abstract}
Domain wall solitons are the simplest topological objects in field theories. 
The conventional translational symmetry in a field theory is the generator of a 
one-parameter family of domain wall solutions, and induces a massless
moduli field which propagates along a domain wall. 
We study similar issues in braided noncommutative field
theories possessing Hopf algebraic translational symmetries.  As a concrete example,
we discuss a domain wall soliton in the scalar $\phi^4$ braided noncommutative field theory in
Lie-algebraic noncommutative spacetime, 
$[x^i,x^j]=2i\kappa \epsilon^{ijk}x_k$ ($i,j,k=1,2,3$),
which has a Hopf algebraic translational symmetry. 
We first discuss the existence of a domain wall soliton 
in view of Derrick's theorem, and construct explicitly a 
one-parameter family of solutions in perturbation of the noncommutativity parameter 
$\kappa$.
We then find the massless moduli field which propagates 
on the domain wall soliton.
We further extend our analysis to the general Hopf algebraic translational symmetry. 
\end{abstract}

\end{titlepage}
\newpage

\section{Introduction}
\renewcommand{\thefootnote}{\arabic{footnote}}
\setcounter{footnote}{0}
Noncommutative field theories \cite{Snyder:1946qz,Yang:1947ud,Connes:1990qp,Doplicher:1994tu} are important subjects for
studying the Planck scale physics. 
The most well-studied are the noncommutative field theories in Moyal spacetime, whose coordinate commutation relation is given by 
$[x^{\mu},x^{\nu}]=i\ct^{\mu\nu}$ with an antisymmetric constant $\ct^{\mu\nu}$. 
Field theories in Moyal spacetime 
are also known to appear as effective field theories of open string theory with a constant background $B_{\mu\nu}$ field \cite{Connes:1997cr,Seiberg:1999vs}. 
Thus, not only as the simplest field theories 
in quantum spacetime but also as toy models of string theory, 
various perturbative and non-perturbative aspects such as unitarity 
\cite{Gomis:2000zz,AlvarezGaume:2001ka,Chu:2002fe}, 
causality \cite{Seiberg:2000gc}, 
UV-IR mixing \cite{Filk:1996dm,Minwalla:1999px,Hayakawa:1999zf},
renormalizability \cite{Minwalla:1999px}, 
scalar solitons \cite{Gopakumar:2000zd,Zhou:2000xg,Hadasz:2001cn,Durhuus:2001nj},
instantons \cite{Nekrasov:1998ss,Furuuchi:1999kv,Aganagic:2000mh,Furuuchi:2000dx,Chu:2001cx,Hamanaka:2001dr}, monopoles \cite{Gross:2000wc,Gross:2000ss,Hamanaka:2001dr}, and other 
solitonic solutions \cite{Polychronakos:2000zm,Jatkar:2000ei,Gross:2000ph,Bak:2000ac,Lechtenfeld:2001aw} have 
extensively been analyzed. 
Recently it has been pointed out 
that Moyal spacetime is invariant under the twisted Poincar\'{e} symmetry, which is a kind of Hopf algebraic symmetry \cite{Chaichian:2004za,Wess:2003da,Koch:2004ud}. There 
have been various proposals to implement the twisted Poincar\'{e} invariance in quantum field theories 
\cite{Oeckl:2000eg,Chaichian:2004yh,Chaichian:2005yp,Balachandran:2005eb,
Balachandran:2005pn,Lizzi:2006xi,
Tureanu:2006pb,Zahn:2006wt,Bu:2006ha,Abe:2006ig,Balachandran:2006pi,Fiore:2007vg,
Joung:2007qv,Sasai:2007me,Riccardi:2007bj}. 
Gravity in fuzzy spacetimes has also been discussed in \cite{Aschieri:2005yw,Aschieri:2005zs,Calmet:2005qm,Kobakhidze:2006kb,Balachandran:2007kv,Mukherjee:2006nd,Kurkcuoglu:2006iw,Banerjee:2007th}.

A prominent feature of Hopf algebraic symmetries
is the general requirement of non-trivial statistics, which is called braiding, of fields to keep the symmetries at quantum level.  
In our previous paper \cite{Sasai:2007me}, it has been
shown that symmetry relations among correlation functions can systematically be derived from Hopf algebraic symmetries in the framework of braided quantum field theories \cite{Oeckl:1999zu}, if appropriate braiding of fields can be chosen.
This feature is in parallel with the existence of similar relations, such as Ward-Takahashi identities, in field theories possessing conventional symmetries.  

The main motivation of this paper is to understand better the physical roles of 
Hopf algebraic symmetries in another setting.
In this paper we study a domain wall soliton in the three-dimensional noncommutative scalar field theory in Lie-algebraic noncommutative space-time $[x^i,x^j]=2i\kappa \epsilon^{ijk} x_k ~(i,j,k=0,1,2)$ \cite{Sasakura:2000vc,Madore:2000en,Imai:2000kq,
Freidel:2005bb,Freidel:2005ec}. This noncommutative spacetime has also a Hopf algebraic Poincar\'{e} symmetry 
\cite{Freidel:2005bb,Freidel:2005ec,Sasai:2007me}, but the difference from 
Moyal spacetime is that 
its translational symmetry is Hopf algebraic, while the rotation-boost symmetry is Hopf algebraic in Moyal spacetime. 
Therefore this noncommutative field theory provides an interesting stage 
for investigating the physical roles of 
the braiding and the Hopf algebraic translational symmetry on a domain wall soliton,
since the conventional translational symmetry in a field theory is the generator of a 
one-parameter family of domain wall solutions, and 
induces a massless moduli field 
which propagates along a domain wall. 

This paper is organized as follows. In section \ref{sec:rev}, we review the three dimensional noncommutative $\phi^4$ theory in the Lie-algebraic noncommutative space-time $[x^i,x^j]=2i\kappa \epsilon^{ijk} x_k$. In section \ref{sec:derrick}, 
we apply the criterion of Derrick's theorem \cite{Derrick:1964ww} 
to the $\phi^4$ theory and conclude that a domain wall solution is possible at least perturbatively in $\kappa$. In section \ref{sec:solh}, 
we solve the equation of motion to obtain a one-parameter family of the kink solutions in perturbation of the noncommutativity parameter $\kappa$. 
In section \ref{sec:modulispace}, we discuss the moduli space.
In section \ref{sec:moduli}, we analyze the moduli field, which propagates along the domain wall soliton, and conclude that the moduli field is massless. 
In section \ref{sec:general}, we study the general Hopf 
algebraic translational symmetry.
The final section is devoted to the summary and comments.

\section{Noncommutative $\phi^4$ theory in Lie-algebraic spacetime 
and the domain wall solutions}
\subsection{Noncommutative $\phi^4$ theory in Lie-algebraic spacetime} \label{sec:rev}
In this subsection, we review the noncommutative $\phi^4$ theory in Lie-algebraic noncommutative space-time whose commutation relation is given by
\begin{equation}
[\xhat^i,\xhat^j]=2i\kappa \epsilon^{ijk} \xhat_k, \label{eq:comx}
\end{equation}
where $i,j,k=0,1,2$\footnote{The signature of the metric $\eta^{ij}$ is $(-1,1,1)$, and that of $\eta^{\mu\nu}$ is $(-1,-1,1,1)$.}\cite{Sasakura:2000vc,Madore:2000en},
following the constructions of 
\cite{Imai:2000kq,Freidel:2005bb,Sasai:2007me}. Imposing the Jacobi identity and Lorentz invariance, we can determine the commutation relations between the 
coordinates and momenta as 
follows \cite{Sasakura:2000vc}:
\begin{equation}
[\Phat^i,\xhat^j]=-i\eta^{ij}\sqrt{1+\kappa^2\Phat^2}+i\kappa\epsilon^{ijk}\Phat_k, \label{eq:comp}
\end{equation}
where we have also imposed $[\Phat^i,\Phat^j]=0$. We can identify these operators with the Lie algebra of $ISO(2,2)$ as follows:
\begin{align}
&\xhat_i=\kappa(\Jhat_{-1,i}-\frac{1}{2}\epsilon_{i}{}^{jk}\Jhat_{jk}), \\
&\Phat_i=\Phat_{\mu=i}, \\
&1+\kappa^2\Phat^{\mu}\Phat_{\mu}=0, \label{eq:hyperboloid}
\end{align}
where the commutation relations of the Lie algebra of $ISO(2,2)$ are given by
\begin{align}
[\Jhat_{\mu\nu},\Jhat_{\rho\sigma}]&=-i(\eta_{\mu\rho}\Jhat_{\nu\sigma}-\eta_{\mu\sigma}\Jhat_{\nu\rho}-\eta_{\nu\rho}\Jhat_{\mu\sigma}+\eta_{\nu\sigma}\Jhat_{\mu\rho}), \\
[\Jhat_{\mu\nu},\Phat_{\rho}]&=-i(\eta_{\mu\rho}\Phat_{\nu}-\eta_{\nu\rho}\Phat_{\mu}), \\
[\Phat_{\mu},\Phat_{\nu}]&=0,
\end{align}
and the Greek indices run through $-1$ to $2$. From the constraint (\ref{eq:hyperboloid}), we can identify the momentum space with the group manifold $SL(2,R)$.

Let $\phi(x)$ be a scalar field in the three-dimensional spacetime. 
Its Fourier transformation is given by
\begin{equation}
\phi(x)=\int dg \tilde{\phi}(g) e^{iP(g)\cdot x},
\end{equation}
where $P^i(g)$ are determined by $g=P^{-1}-i\kappa P^i\tilde{\sigma}_i\in SL(2,R)$\footnote{The definition of $\tilde{\sigma}_i$ is given by \[\tilde{\sigma}_0=
\sigma_2, ~\tilde{\sigma}_1=i\sigma_3,~ \tilde{\sigma}_2=i\sigma_1,\] with Pauli 
matrices ($\sigma_1, \sigma_2,\sigma_3$). We have also changed the normalization of 
$P^{-1}$ by $\kappa$ from (\ref{eq:hyperboloid}).} and $\int dg$ is the Haar measure 
of $SL(2,R)$. This $P^{-1}$ can take two values,
\begin{equation}
P^{-1}=\pm\sqrt{1+\kappa^2P^iP_i}, \label{eq:defofp-1}
\end{equation}
for each $P_i$. This unphysical two-fold degeneracy can be deleted by
imposing
\begin{equation}
\tilde{\phi}(g)=\tilde{\phi}(-g).
\end{equation}

The definition of the star product is given by\footnote{In fact, one can produce the commutation relation between the coordinates (\ref{eq:comx}) 
by differentiating both hand sides of (\ref{eq:defofstar}) with respect to 
$P_1^i\equiv P^i(g_1)$ and  $P_2^j\equiv P^j(g_2)$ and then taking the limit $P_1^i, P_2^i\to 0$.}
\begin{equation}
e^{iP(g_1)\cdot x}\star e^{iP(g_2)\cdot x}=e^{iP(g_1g_2)\cdot x}. \label{eq:defofstar}
\end{equation}
This determines the coproducts of $P^i$ and $P^{-1}$ via the group product $g_1g_2$ as 
\begin{align}
\Delta P^i&=P^i\otimes P^{-1}+P^{-1}\otimes P^i-\kp \epsilon^{ijk}P_j\otimes P_k, \label{eq:coprop} \\
\Delta P^{-1}&=P^{-1}\otimes P^{-1}+\kp^2 P^i\otimes P_i. \label{eq:coprop-1}
\end{align}

We consider the $\phi^4$ theory in the Lie-algebraic noncommutative space-time. We give the action as follows:
\begin{equation}
S=\int d^3x\bigg[-\frac{1}{2}(\partial^i\phi \star \partial_i\phi)(x)+\frac{1}{2}m^2(\phi \star \phi)(x)-\frac{\lambda}{4}(\phi \star \phi \star \phi \star \phi)(x)-\frac{m^4}{4\lambda} \bigg], \label{eq:action}
\end{equation}
where we have chosen the constant term so that the minima of the potential vanish
when $\kappa =0$. 

Carrying out the coordinate integration, one finds a modified energy-momentum conservation: $P^i(g_1g_2\cdots )=P_1+P_2+\cdots +\mathcal{O}(\kp)=0$ 
at the classical level. 
This should be regarded as a consequence of the Hopf algebraic translational symmetry.
A naive construction of noncommutative quantum field theory in this space-time leads to disastrous violations of the energy-momentum conservation in the non-planar diagrams \cite{Imai:2000kq}. One can avoid this violation by
introducing a nontrivial statistics between scalar fields, which is given by
\begin{equation}
\psi(\tilde{\phi}_1(g_1)\tilde{\phi}_2(g_2))=\tilde{\phi}_2(g_2)\tilde{\phi}_1(g_2^{-1}g_1g_2), \label{eq:braid}
\end{equation}
where $\psi$ is an exchanging map. This is denoted by braiding. This braiding was
first derived from three dimensional quantum gravity with scalar particles \cite{Freidel:2005bb}. 
With this braiding, correlation functions respect the Hopf algebraic symmetry 
at the full quantum level \cite{Sasai:2007me}. 

\subsection{Derrick's theorem in the noncommutative $\phi^4$ theory} \label{sec:derrick}
We consider a domain wall soliton in the noncommutative $\phi^4$ theory. At first we consider whether the domain wall solution may exist or not by applying
the criterion of Derrick's theorem
\cite{Derrick:1964ww}.

Varying the action (\ref{eq:action}) with respect to $\phi(x)$, we obtain the equation of motion,
\begin{equation}
\partial^2\phi(x)+m^2\phi(x)-\lambda (\phi\star \phi \star \phi)(x)=0. \label{eq:eom}
\end{equation}
Since our interest is in a domain wall, we consider only one spatial direction
of the coordinates\footnote{In one dimension, there is no 
non-trivial noncommutativity of coordinates, but the coordinate and the momentum are noncommutative as in (\ref{eq:comp}). Thus a soliton solution is not the same as the commutative case.}. Let us change the variables $P, P^{-1}$ as 
follows\footnote{When one considers only spatial directions, one can safely take only the positive branch of $P^{-1}$ in (\ref{eq:defofp-1}).}:
\begin{equation}
\label{eq:ptheta}
P=\frac{1}{\kappa}\sinh (\kappa \theta) ~~~~P^{-1}=\cosh (\kappa \theta),
\end{equation}
where $-\infty < \ct <\infty $. Then the field $\phi(x)$ is given by
\begin{equation}
\phi(x)=\int \frac{d \theta}{2\pi} \tilde{\phi}(\theta)e^{\frac{i}{\kappa}\sinh (\kappa \theta)x}. \label{eq:1dphi}
\end{equation}
The star product simply becomes
\begin{equation}
e^{\frac{i}{\kappa}\sinh (\kappa \theta_1)x}\star e^{\frac{i}{\kappa}\sinh (\kappa \theta_2)x}=e^{\frac{i}{\kappa}\sinh (\kappa (\theta_1+\theta_2))x}. \label{eq:1dstar}
\end{equation}

Here we notice that the nontrivial momentum sum, which comes from the star product, can be described by the usual sum of $\theta$. In fact, from (\ref{eq:comp}), we can find that the commutation relation between $\hat{\theta}=\frac{1}\kappa \sinh^{-1}(\kappa
\hat{P})$ and $\hat{x}$ becomes
\begin{equation}
[\hat{\theta},\hat{x}]=-i, \label{eq:comc}
\end{equation}
and, from (\ref{eq:coprop}), the coproduct of $\hat{\theta}$ becomes
\begin{equation}
\label{eq:usualrule}
\Delta \hat{\theta} =\hat{\theta} \otimes 1+1\otimes \hat{\theta},
\end{equation}
which is the usual Leibnitz rule. 

Using (\ref{eq:1dphi}) and (\ref{eq:1dstar}), the equation of motion (\ref{eq:eom}) becomes
\begin{align}
&\int \frac{d\theta}{2\pi }
\bigg(
-\frac{1}{\kappa^2}\sinh^2(\kappa \theta)\ftilde(\theta)+m^2 \ftilde(\theta) \notag \\
&-\lambda \int \frac{d\theta_1}{2\pi }\frac{d\theta_2}{2\pi }\frac{d\theta_3}{2\pi }(2\pi)\delta (\theta-\theta_1-\theta_2-\theta_3)\ftilde(\ct_1)\ftilde(\ct_2)\ftilde(\ct_3)
\bigg)
e^{\frac{i}{\kappa}\sinh(\kappa \theta)x}=0.
\end{align}
Thus we find that
\begin{equation}
\bigg(-\frac{1}{\kappa^2}\sinh^2 (\kappa \theta)+m^2\bigg)\frac{\tilde{\phi}(\theta)}{2\pi}-\lambda \int \frac{d\theta_1}{2\pi}\frac{d\theta_2}{2\pi}\frac{d\theta_3}{2\pi}\delta (\theta -\theta_1-\theta_2-\theta_3)\ftilde(\ct_1)\ftilde(\ct_2)\ftilde(\ct_3)=0. \label{eq:eomct}
\end{equation}

Next we define
\begin{equation}
\label{eq:hphitheta}
h(x)=\int \frac{d\ct}{2\pi}\ftilde(\ct)e^{i\ct x}.
\end{equation}
Rewriting (\ref{eq:eomct}) with $h(x)$, we obtain an equation of motion for $h(x)$:
\begin{equation}
\frac{1}{\kappa^2}\sin^2(\kp \del)h(x)+m^2h(x)-\lambda h^3(x)=0. \label{eq:1deom}
\end{equation}
Now the equation has a familiar local interaction term, but has
infinite higher derivative terms. Another very important feature is that, though the star product (\ref{eq:1dstar}) and hence (\ref{eq:eom}) are not invariant 
under the simple translation 
$x\rightarrow x+a$, the equation (\ref{eq:1deom}) has the obvious translational
symmetry. 

To analyze (\ref{eq:1deom}), we may consider an action for $h(x)$, which is given by
\begin{equation}
S_h=\int dx \bigg[-\frac{1}{2\kp^2}\sin (\kp \del)h(x)\sin (\kp \del)h(x)+\frac{1}{2}m^2h^2(x)-\frac{\lambda}{4}h^4(x)-\frac{m^4}{4\lambda}\bigg]. 
\end{equation}
Then the problem becomes to find the minimum of the energy $E_h=-S_h$ with
an appropriate boundary condition at the infinities $x\rightarrow \pm \infty$,
where the field takes the vacuum values $h=\pm m/\sqrt{\lambda}$.  

In this regard,
we will consider perturbation in $\kappa$.
The energy can be expanded in the form,
\begin{align}
E_h=-S_h=\int dx \left[\frac{1}{2}\left( \sum_{n=1}^\infty \kappa^{2n-2} C_n \, \partial^n h(x) \partial^n h(x)
\right) +V\left(h(x)\right)\right],
\end{align}
where $C_n=2^{n-1}/(n!(2n-1)!!)$ and  $V\left(h(x)\right)=-\frac{1}{2}m^2h^2(x)+\frac{\lambda}{4}h^4(x)+\frac{m^4}{4\lambda} 
\geq 0$.
The positivity of all the coefficients $C_n$ will play an essential role in the following discussions.

Let us rescale $x^i\to x^{'i}=\mu\, x^i$ ($0<\mu<\infty$) and define $h^{(\mu)}(x)=h(\mu x)$. Derrick's theorem \cite{Derrick:1964ww} tells us that if the energy for the rescaled field does not have any stationary points with respect to $\mu$, there exist no soliton solutions. In our case, the energy for $h^{(\mu)}(x)$ is given by
\begin{align}
E_{h^{(\mu)}}&=\int dx \left[\frac{1}{2}\left( \sum_{n=1}^{\infty} \kappa^{2n-2} C_n \, \partial^n h^{(\mu)}(x) \partial^n h^{(\mu)}(x)
\right) +V\left(h^{(\mu)}(x)\right)\right] \\
&=\int dx'\frac{1}{\mu} \left[\frac{1}{2}\left( \sum_{n=1}^{\infty} \mu^{2n}\kappa^{2n-2} C_n \, \partial^{'n} h(x') \partial^{'n} h^{(\mu)}(x')
\right) +V\left(h^{(\mu)}(x')\right)\right] \\
&=\frac{1}{\mu}E_0+\sum_{n=1}^{\infty} \mu^{2n-1} E_{2n},  \label{eq:rescaledenergy}
\end{align}
where 
\begin{align}
E_0&=\int dx\, V\left(h(x)\right), \notag \\
E_{2n}&=\frac{C_n}{2} \int dx\, (\del^{n} h(x))^2.
\end{align}
All the $E_0$ and $E_{2n}$ are non-negative in general.  
For an $h(x)$ connecting the distinct vacua,
$E_0$ and at least some of the $E_{2n}$ are positive. Therefore 
(\ref{eq:rescaledenergy}) diverges at 
$\mu\rightarrow +0,+\infty$(or a finite $\mu_c$)\footnote{For example,
the convergence radius of the infinite sum is 
$|\mu|<\mu_c=\pi/4\kappa$  for $h(x)=\tanh (x)$.},
and takes a minimum value at a positive finite $\mu$. Thus we
conclude that a domain wall solution in this noncommutative field theory is possible.

\subsection{The perturbative solution of $h(x)$} \label{sec:solh}
Next we consider the perturbative solution of $h(x)$. We write the perturbation series as $h(x)=h_0(x)+\kp^2h_2(x)+\kp^4h_4(x)+\cdots $. 
Inserting this into the equation of motion (\ref{eq:1deom}), 
we obtain for each order of $\kp^2$,
\begin{align}
&\del^2h_0(x)+2h_0(x)-2h_0^3(x)=0, \label{eq:0ordereq} \\
&\del^2h_2(x)+2h_2(x)-6h_0^2(x)h_2(x)-\frac{1}{3}\del^4h_0(x)=0, \label{eq:2ordereq} \\
&\del^2h_4(x)+2h_4(x)-6h_0^2(x)h_4(x)-\frac{1}{3}\del^4h_2(x)+\frac{2}{45}\del^6h_0(x)-6h_0(x)h_2^2(x)=0, \label{eq:4ordereq} \\
&~~~~~~~~~~~~~~~~~~~~~\vdots \notag
\end{align}
where we have set $m^2=2, \lambda=2$ for simplicity.

Our purpose is to obtain kink solutions whose boundary condition is given by $h(x=\pm \infty)=\pm 1$.
The equation (\ref{eq:0ordereq}) is the same as the equation of motion 
in the commutative case. The general solution of (\ref{eq:0ordereq}) has two 
integration constants. One is interpreted as the translation of the solution, and the other can be determined by the behavior at $x=- \infty $ 
or $\infty$. If one assumes $h_0(x=\pm \infty)\neq \pm 1$, $h_0(x)$ diverges or oscillates at $x=\pm\infty$. For such an $h_0(x)$,
the solutions of $h_{2n}(x)\ (n=1,2,\cdots)$ diverge 
at $x=\pm \infty$, unless $h_{2n}(x=\pm \infty)=0$. Thus the boundary condition 
$h(x=\pm \infty)=\pm 1$ cannot be satisfied by the perturbative solution, unless we
assume $h_0(x=\pm \infty)= \pm 1$. 

For the boundary condition $h_0(x=\pm \infty)= \pm 1$, 
the solution to the equation (\ref{eq:0ordereq}) is 
well known and given by
\begin{equation}
\label{eq:tanh}
h_0(x)=\tanh (x+a),
\end{equation}
where $a\in \mathbb{R}$. The arbitrary parameter $a$ results from the translational
 invariance of the equation (\ref{eq:0ordereq}).

Next we will solve the equation (\ref{eq:2ordereq}) for $a=0$. Let us put
\begin{equation}
h_2(x)=\frac{f(x)}{\cosh^2x}. \label{eq:defofh2}
\end{equation}
Inserting this and (\ref{eq:tanh}) for $a=0$ into (\ref{eq:2ordereq}), we obtain
\begin{equation}
f''(x)-4\tanh xf'(x)-\frac{8}{3}\bigg(2\frac{\tanh x}{\cosh^2x}-\tanh^3x\bigg)=0. \label{eq:feq}
\end{equation}
Then let us put 
\begin{equation}
f'(x)=\cosh^4(x)\,g(x),
\end{equation}
and insert this into (\ref{eq:feq}). The equation becomes
\begin{equation}
g'(x)=\frac{8}{3\cosh^4x}\bigg(2\frac{\tanh x}{\cosh^2x}-\tanh^3x\bigg). \label{eq:geq}
\end{equation}
Integrating (\ref{eq:geq}) over $x$, we obtain
\begin{equation}
g(x)=\frac{2}{3\cosh^4x}-\frac{4}{3\cosh^6x}+A_1,
\end{equation}
where $A_1$ is an integration constant. 
Thus the differential equation of $f(x)$ becomes
\begin{equation}
f'(x)=\frac{2}{3}-\frac{4}{3\cosh^2x}+A_1\cosh^4x.
\end{equation}
Integrating this over $x$ and using (\ref{eq:defofh2}), we obtain
\begin{equation}
h_2(x)=\frac{2x}{3\cosh^2x}-\frac{4\tanh x}{3\cosh^2x}+A_1\bigg(\frac{3x}{8\cosh^2x}+\frac{3}{8}\tanh x+\frac{1}{4}\cosh^2x\tanh x\bigg)+\frac{A_2}{\cosh^2x},
\end{equation}
where $A_2$ is an integration constant. 

Since the term with $A_1$ is divergent at 
$x=\pm\infty$, we have to put $A_1=0$ from the boundary condition.  
The $A_2$ term is allowed but can just be 
absorbed into the parameter $a$ in (\ref{eq:tanh}), because 
$\tanh(x+\kappa^2 A_2)=\tanh(x)+\kappa^2 A_2 /\cosh^2(x)+\cdots$.
To systematically kill such redundant integration constants, we impose the 
oddness condition,
$h_{2n}(x)=-h_{2n}(-x)$ for $a=0$. Then $A_2=0$ is also required. 
Finally, recovering the parameter $a$, we obtain 
\begin{equation}
h_2(x)=\frac{2(x+a)}{3\cosh^2(x+a)}-\frac{4\tanh (x+a)}{3\cosh^2(x+a)}.
\end{equation}

In the same way, we can obtain the solution to the equation (\ref{eq:4ordereq}),
which is given by 
\begin{align}
h_4(x)&=\frac{134(x+a)}{45\cosh^2(x+a)}-\frac{8(x+a)}{3\cosh^4(x+a)}-\frac{40\tanh (x+a)}{9\cosh^2(x+a)} \notag \\
&-\frac{4(x+a)^2\tanh (x+a)}{9\cosh^2(x+a)}+\frac{52\tanh (x+a)}{9\cosh^4(x+a)}.
\end{align}
This procedure will be able to be repeated to a required order.

\subsection{The solution of $\phi(x)$ and the moduli space}
\label{sec:modulispace}
In the preceding subsection, we have obtained the perturbative 
solution of $h(x)$. Then we formally know the perturbative soliton solution of
$\phi(x)$ through $\ftilde(\theta)$, which are related to $\phi(x)$ and $h(x)$  
by (\ref{eq:1dphi}) and (\ref{eq:hphitheta}), respectively. 

In the following let us discuss the moduli space of the domain wall solution. 
In $h(x)$, the moduli parameter is just the translation parameter $a$. 
This translation corresponds to 
the phase rotation $\ftilde(\theta)\rightarrow e^{ia\theta}\ftilde(\theta)$, as
can be seen in (\ref{eq:hphitheta}).
Therefore the translation on $\phi(x)$ is given by
\begin{align}
T_a \phi(x)
&=\int \frac{d\ct}{2\pi }\ftilde(\theta)e^{i(\ct a+\frac{1}{\kappa}\sinh (\kappa \ct)x)} \\
&=e^{ia\hat{\theta}} \phi(x). \label{eq:solofphia}
\end{align}
This last expression shows that the operator $\hat{\theta}$, which is a
non-linear function of 
$\hat P$, is the generator of the translational moduli. 
In fact, by using the Leibnitz rule (\ref{eq:usualrule}) and following the 
same procedure as a conventional symmetry,
one can directly show that, if $\phi(x)$ is a solution to the equation of motion 
(\ref{eq:eom}), $e^{ia\hat{\theta}} \phi(x)$ is also a solution.
The generalization of this fact to the general Hopf algebraic translational
symmetry will be discussed in section \ref{sec:general}.

\subsection{The moduli field from the Hopf algebraic translational \\ symmetry} 
\label{sec:moduli}
Another interesting consequence of the conventional translational symmetry in a field
theory is the existence of a massless propagating field along a domain wall. 
This field can be obtained by generalizing the constant moduli parameter $a$ 
to a field $a(x_\|)$ depending on the coordinates along a domain wall.  
In this subsection, we will study this aspect in our noncommutative field theory.

We go back to the three dimensional case. For simplicity, we set $\kappa =1$. We change the variable $P^i(g)$ as follows:
\begin{equation}
P^i=\sinh (\sqrt{k^2})\frac{k^i}{\sqrt{k^2}}.
\end{equation}
This $k^i$ is the three-dimensional analog of $\theta$ in the previous subsections.
The field $\phi(x)$ can be rewritten as 
\begin{align}
\phi(x)&=\int \frac{d^3P}{(2\pi)^3\sqrt{1+P^2}}\tilde{\phi}(P)e^{iP\cdot x} \notag \\
&=\int \frac{d^3k}{(2\pi)^3} \frac{\sinh^2 (\sqrt{k^2})}{k^2}\tilde{\phi}(k)e^{i\sinh (\sqrt{k^2})\frac{k^i}{\sqrt{k^2}}x_i} \notag \\
&\equiv\int \frac{d^3k}{(2\pi)^3} \tilde{\varphi}(k)e^{i\sinh (\sqrt{k^2})\frac{k^i}{\sqrt{k^2}}x_i}.
\end{align}
Let us define
\begin{equation}
h(\hat{x})=\int \frac{d^3k}{(2\pi)^3} \tilde{\varphi}(k)e^{ik\cdot \hat{x}}
\end{equation}
as in (\ref{eq:hphitheta}). Then it can be shown that
the action (\ref{eq:action}) is equivalent to the following action \cite{Sasakura:2000vc}:
\begin{equation}
\label{eq:hathaction}
S=\langle 0|\left(-\frac{1}{2}h(\xhat)[\Phat^i,[\Phat_i,h(\xhat)]]+\frac{1}{2}m^2h(\xhat)^2-\frac{\lambda}{4}h(\xhat)^4 \right)|0\rangle,
\end{equation}
where $|0\rangle$ denotes the momentum zero eigenstate $\Phat^i|0\rangle =0$, and
\begin{align}
[\Phat^i,\xhat^j]&=-i\eta^{ij}\sqrt{1+\Phat^2}+i\epsilon^{ijk}\Phat_k, \label{eq:comrelPx} \\
[\Phat^i,\Phat^j]&=0.
\end{align}
From the commutation relation, the following relation is satisfied \cite{Sasakura:2000vc}:
\begin{align}
\Phat^ie^{ik\cdot \xhat}|0\rangle &=\sinh (\sqrt{k^2})\frac{k^i}{\sqrt{k^2}}e^{ik\cdot \xhat}|0\rangle \notag \\
&=P^ie^{ik\cdot \xhat}|0\rangle.
\end{align}
Thus $e^{ik\cdot \xhat}|0\rangle $ is the eigenstate of $\Phat^i$ 
with an eigenvalue $P^i$. In the following discussions, we use the notation $|P^i\rangle \equiv e^{ik\cdot \xhat}|0\rangle $.

The equation of motion from (\ref{eq:hathaction}) is 
\begin{equation}
\left(-[\Phat^2,h(\xhat)]+m^2 h(\xhat)-\lambda h(\xhat)^3\right)|0\rangle =0. 
\label{eq:operatoreom}
\end{equation}
As has been discussed in the preceding subsections, there exists a one-parameter family
of domain wall solutions $h_{sol}^a(\hat x)$ to (\ref{eq:operatoreom}), where
$a$ is the translational parameter. 
One may expand the solution with respect to $a$ as 
$h_{sol}^a(\xhat)=h_{sol}(\xhat^{1})+af(\xhat^1)+\cdots$, where
we have chosen $\hat x^1$ as the spatial direction perpendicular to the domain 
wall\footnote{The following discussions do not depend on the value of $a$ 
where the expansion with respect to $a$ is carried out.}.
Then, putting this expansion into (\ref{eq:operatoreom}) and 
taking the first order of $a$, $f(\xhat^1)$ is shown to satisfy
\begin{equation}
\left(-[\Phat^2,f(\xhat^1)]+m^2f(\xhat^1)-3\lambda (h_{sol}(\xhat^1))^2f(\xhat^1)\right)|0\rangle=0. \label{eq:feq2}
\end{equation}

To study the property of the moduli field, we will replace $a$ to $a(\hat{x_0}, \hat{x_2})$. In doing so, the braiding property (\ref{eq:braid}) plays essential roles. For general 
$h_1(\hat x), h_2(\hat x)$, we have the following commuting property,
\begin{align}
h_1(\xhat)h_2(\xhat)&=\int dg_1\int dg_2 ~\ftilde_1(g_1)\ftilde_2(g_2)e^{ik(g_1)\cdot \xhat}e^{ik(g_2)\cdot \xhat} \notag \\
&=\int dg_1\int dg_2 ~\ftilde_2(g_2)\ftilde_1(g_2^{-1}g_1g_2)e^{ik(g_1)\cdot \xhat}e^{ik(g_2)\cdot \xhat} \notag \\
&=\int dg_1\int dg_2 ~\ftilde_2(g_2)\ftilde_1(g_1)e^{ik(g_2g_1g_2^{-1})\cdot \xhat}e^{ik(g_2)\cdot \xhat} \notag \\
&=\int dg_1\int dg_2 ~\ftilde_2(g_2)\ftilde_1(g_1)e^{ik(g_2g_1)\cdot \xhat} \notag \\
&=h_2(\xhat)h_1(\xhat), \label{eq:braidprop}
\end{align}
where we have used the invariance of the Haar measure. 
Inserting $h(\xhat)=h_{sol}(\xhat^{1})+a(\hat{x_0}, \hat{x_2})f(\xhat^1)$ 
into the equation of motion (\ref{eq:operatoreom}) and taking the first order 
of $a(\hat{x_0}, \hat{x_2})$, we obtain
\begin{equation}
\left(-[\Phat^2,a(\hat{x_0}, \hat{x_2})f(\xhat^1)]+m^2a(\hat{x_0}, \hat{x_2})f(\xhat^1) -3\lambda a(\hat{x_0}, \hat{x_2})(h_{sol}(\xhat^{1}))^2f(\xhat^1)\right)|0\rangle=0.
\end{equation}
Then, from (\ref{eq:feq2}), we obtain
\begin{equation}
[\Phat^2,a(\hat{x_0}, \hat{x_2})]f(\xhat^1)|0\rangle=0.
\end{equation}
After the Fourier transformation, we find
\begin{equation}
\int _{P_1}\int_{P_2}\tilde{a}(P_1)\tilde{f}(P_2)(P(g_1g_2)^2-P_2^2)|P(g_1g_2)\rangle =0, \label{eq:modulieq1}
\end{equation}
where $P_1^i=(P_1^0,0,P_1^2)$ and $P_2^i=(0,P_2,0)$.  From the formula of the coproduct of $P^2$, which is given by
\begin{align}
\Delta (P^2)&=P^2\otimes 1+1\otimes P^2 + P^2\otimes P^2 \notag \\
&+2\sqrt{1+P^2}P^i\otimes \sqrt{1+P^2}P_i+P^iP^j\otimes P_iP_j, \label{eq:p2copro}
\end{align}
the equation (\ref{eq:modulieq1}) becomes
\begin{equation}
\int _{P_1}\int_{P_2}P_1^2\tilde{a}(P_1)(1+P_2^2)\tilde{f}(P_2)|P(g_1g_2)\rangle =0.
\end{equation}
Operating $\langle x|$ from the left, we find
\begin{align}
&\int _{P_1}\int_{P_2}P_1^2\tilde{a}(P_1)(1+P_2^2)\tilde{f}(P_2)e^{iP(g_1g_2)\cdot x} \notag \\
=&\int _{P_1}\int_{P_2}P_1^2\tilde{a}(P_1)(1+P_2^2)\tilde{f}(P_2)e^{iP_1\cdot x}\star e^{iP_2\cdot x} \notag \\
=&-\del^2 a(x_0,x_2)(1-\del^2)f(x_1) \notag \\
=&0.
\end{align}
Thus, since $(1-\del^2)f(x_1)$ does not vanish constantly, we obtain
\begin{equation}
\del^2 a(x_0,x_2)=0.
\end{equation}
Thus we conclude that the moduli field is massless.

The preceding discussions in the operator formalism can be repeated 
with the star product. Putting the expansion $\phi_{sol}^a(x)=\phi_{sol}(x^1)+a\, g(x^1)+\cdots$ into the equation of motion (\ref{eq:eom}), one obtains
\begin{equation}
\del^2 g(x^1)+m^2 g(x^1)-3\lambda \f_{sol}(x^1)\star \f_{sol}(x^1)\star g(x^1)=0.
\label{deltafeom}
\end{equation}
Next we define the moduli field $a(x^0,x^2)$, and consider $\phi(x)=\phi_{sol}(x^1)+a(x^0,x^2)\star g(x^1)$. 
Putting this into the equation of motion and taking the first order of $a(x^0,x^2)$,
we obtain
\begin{equation}
\del^2(a(x^0,x^2)\star g(x^1))+m^2(a(x^0,x^2)\star g(x^1))-3\lambda \f_{sol}(x^1)\star \f_{sol}(x^1)\star a(x^0,x^2)\star g(x^1)=0, \label{eq:aeom}
\end{equation}
where we have used the property similar to (\ref{eq:braid}) for the star product. The first term of (\ref{eq:aeom}) can easily be computed by using the coproduct of $P^{2}$. Using (\ref{eq:p2copro}) and (\ref{eq:defofp-1}), $\del^2 (a\star g)$ becomes
\begin{equation}
\del^2 (a(x^0,x^2)\star g(x^1))=a(x^0,x^2)\star \del^2 g(x^1)+\del^2a(x^0,x^2)
\star g(x^1)-\del^2a(x^0,x^2)\star \del^2g(x^1).
\end{equation}
Thus (\ref{eq:aeom}) becomes
\begin{align}
0&=a(x^0,x^2)\star 
(\del^2 g(x^1)+m^2g(x^1)-3\lambda g(x^1)\star \f_{sol}(x)\star \f_{sol}(x)) \notag \\
&\ \ \ +\del^2a(x^0,x^2)\star g(x^1)-\del^2a(x^0,x^2)\star \del^2g(x^1) \notag \\
&=(g(x^1)-\del^2g(x^1))\star \del^2a(x^0,x^2), \label{eq:aeom1}
\end{align}
where we have used (\ref{deltafeom}). Thus we obtain the same conclusion as above.

\section{~The general Hopf algebraic translational \\ ~symmetry}
\label{sec:general}
In the preceding section, the discussions are restricted to the specific noncommutative field theory. However, it is interesting to know what holds for 
the general Hopf algebraic translational symmetry. In this section, we will show 
that the results in the preceding section 
are the general consequence of a Hopf algebraic translational symmetry. 

We first assume that, in considering domain wall solutions, only one direction of 
momentum is relevant. Then the (associative) coproduct of the momentum may be 
written as
\begin{equation}
\Delta(\hat P)=\sum_i a_i(\hat P)\otimes b_i(\hat P).
\end{equation}
This defines the associative sum of two momenta $\oplus$.

Let us consider a small momentum $P_\varepsilon$. One may consider its $n$ sum,
\begin{equation}
P_n\equiv \overbrace{P_\varepsilon\oplus P_\varepsilon\oplus\cdots\oplus 
P_\varepsilon}^n.
\end{equation}
For such $P_n$, let us define
\begin{equation}
\theta(P_n)=n P_\varepsilon.
\end{equation}
Then $\theta(P)$ can be shown to define an additive quantity for $\oplus$ as
\begin{align}
\theta(P_n\oplus P_m)&=\theta(P_{n+m})\\
&=(n+m)P_\varepsilon \\
&=\theta(P_n)+\theta(P_m),
\end{align}
where we have used the associativity of $\oplus$.
This shows the usual Leibnitz rule for the coproduct of $\hat\theta$,
\begin{equation}
\label{eq:usualtheta}
\Delta(\hat \theta)=\hat \theta \otimes 1+1\otimes \hat \theta.
\end{equation} 
The above discussions may be generalized to negative $n$'s, 
and further to a continuous momentum by considering 
the limit $P_\varepsilon\rightarrow 0$.

In the actual computation, it is convenient to consider a differential equation for $\theta(P)$ as
\begin{equation}
\label{eq:diffeqtheta}
\frac{d \theta(P)}{d P}=\lim_{P_\varepsilon\rightarrow 0} 
\frac{\theta(P_\varepsilon\oplus P)-\theta(P)}{P_\varepsilon\oplus P-P}=
\lim_{P_\varepsilon\rightarrow 0} 
\frac{P_\varepsilon}{P_\varepsilon\oplus P-P}.
\end{equation}
The last limit can be computed from a given coproduct of 
momentum\footnote{For the limit to have a finite value, $0\oplus P=P$ is necessary.
This is mathematically obtained from the axiom 
$({\rm id}\otimes \epsilon)\Delta=(\epsilon\otimes {\rm id})\Delta=1$
with $\epsilon(P)=0$, where $\epsilon$ is the counit map.}. 
The initial condition should be taken as $\theta(0)=0$.
 
For the noncommutative field theory in the preceding section, the coproduct of
momentum is given by (\ref{eq:coprop}), and the differential equation 
(\ref{eq:diffeqtheta}) becomes
\begin{equation}
\frac{d \theta(P)}{d P}=\lim_{P_\varepsilon\rightarrow 0} 
\frac{P_\varepsilon}{\sqrt{1+\kappa^2 P^2}P_\varepsilon+\sqrt{1+\kappa^2 
P_\varepsilon^2}P -P}
=\frac{1}{\sqrt{1+\kappa^2 P^2}}.
\end{equation}
With the initial condition $\theta(0)=0$, 
the solution is actually given by $P=\frac{1}{\kappa}\sinh(\kappa \theta)$,
which agrees with (\ref{eq:ptheta}).

As explained in section \ref{sec:moduli}, the usual Leibnitz rule 
(\ref{eq:usualtheta}) for $\hat\theta$ implies 
that $e^{ia\hat\theta} \phi(x)$ forms a one-parameter family of domain wall solutions, provided that $\phi(x)$ is such a solution. 
It would be physically reasonable to assume that there exists at least one domain 
wall solution which connects distinct vacua with the same energy, 
if a theory has multiple vacua and is physically sensible.  
Therefore a noncommutative field theory 
possessing a Hopf algebraic translational symmetry will have a one-parameter family
of domain wall solutions, if it has multiple vacua with the same energy. 
The associated moduli field
will also have a vanishing mass, since the zero mode of the moduli field is 
the parameter itself, and its potential should be flat in this direction.

\section{Summary and comments}
We have studied the domain wall soliton and its moduli field in the 
braided $\phi^4$ noncommutative field theory in the three dimensional Lie algebraic 
noncommutative spacetime $[x^i,x^j]=2i\kappa \epsilon^{ijk}x_k$. 
This noncommutative spacetime is known to have a Hopf algebraic translational
symmetry, and provides an interesting stage for investigating the physical roles 
of a Hopf algebraic translational symmetry on domain walls. 
We have found that there exists a one-parameter family of the solutions, 
and the mass of the moduli field propagating along the domain wall vanishes.
We have also argued that the results should also hold 
in the general noncommutative field theory with a Hopf algebraic translational 
symmetry.  
This conclusion agrees with what can be obtained from the conventional translational symmetry 
of the usual field theory. 
Therefore our results show another evidence for the physical importance of 
Hopf algebraic symmetries as much as the standard Lie-algebraic symmetries.  

Two comments are in order. 
Firstly, we have used the braiding property when we analyze 
the equation of motion for the moduli field. Therefore the non-trivial 
statistics of both the domain wall and the moduli field seem to play
significant roles in their dynamics.  
Secondly, in our discussions, the operator $\hat\theta$, which has a Lie-algebraic 
coproduct, plays essential roles. 
The general derivation of $\hat\theta$ in the preceding section 
is based on that there is only one relevant direction. Therefore, if one 
considers higher dimensional topological objects such as instantons,
it is not at all clear whether we will obtain results in parallel with the usual field theories.

\section*{Acknowledgments}

We would like to thank R.~Sasaki for useful discussions and comments.
Y.S. was supported in part by JSPS Research Fellowships for Young Scientists.
N.S. was supported in part by the Grant-in-Aid for Scientific Research Nos. 16540244 and 18340061
from the Ministry of Education, Science, Sports and Culture of Japan.


\newpage
\begin{thebibliography}{10}
\bibitem{Snyder:1946qz}
  H.~S.~Snyder,
  ``Quantized space-time,''
  Phys.\ Rev.\  {\bf 71}, 38 (1947).


\bibitem{Yang:1947ud}
  C.~N.~Yang,
  ``On Quantized Space-Time,''
  Phys.\ Rev.\  {\bf 72}, 874 (1947).

\bibitem{Connes:1990qp}
  A.~Connes and J.~Lott,
  ``Particle Models And Noncommutative Geometry (Expanded Version),''
  Nucl.\ Phys.\ Proc.\ Suppl.\  {\bf 18B}, 29 (1991).

\bibitem{Doplicher:1994tu}
  S.~Doplicher, K.~Fredenhagen and J.~E.~Roberts,
  ``The Quantum structure of space-time at the Planck scale and quantum fields,''
  Commun.\ Math.\ Phys.\  {\bf 172}, 187 (1995)
  [arXiv:hep-th/0303037].

\bibitem{Connes:1997cr}
  A.~Connes, M.~R.~Douglas and A.~S.~Schwarz,
  ``Noncommutative geometry and matrix theory: Compactification on tori,''
  JHEP {\bf 9802}, 003 (1998)
  [arXiv:hep-th/9711162].
  
\bibitem{Seiberg:1999vs}
  N.~Seiberg and E.~Witten,
  ``String theory and noncommutative geometry,''
  JHEP {\bf 9909}, 032 (1999)
  [arXiv:hep-th/9908142].


\bibitem{Gomis:2000zz}
J.~Gomis and T.~Mehen,
``Space-time noncommutative field theories and unitarity,''
Nucl.\ Phys.\ B {\bf 591}, 265 (2000)
[arXiv:hep-th/0005129].

\bibitem{AlvarezGaume:2001ka}
  L.~Alvarez-Gaume, J.~L.~F.~Barbon and R.~Zwicky,
  ``Remarks on time-space noncommutative field theories,''
  JHEP {\bf 0105}, 057 (2001)
  [arXiv:hep-th/0103069].
  
\bibitem{Chu:2002fe}
  C.~S.~Chu, J.~Lukierski and W.~J.~Zakrzewski,
  ``Hermitian analyticity, IR/UV mixing and unitarity of noncommutative  field
  theories,''
  Nucl.\ Phys.\ B {\bf 632}, 219 (2002)
  [arXiv:hep-th/0201144].
  
\bibitem{Seiberg:2000gc}
N.~Seiberg, L.~Susskind and N.~Toumbas,
``Space/time non-commutativity and causality,''
JHEP {\bf 0006}, 044 (2000)
[arXiv:hep-th/0005015].

\bibitem{Filk:1996dm}
  T.~Filk,
  ``Divergencies in a field theory on quantum space,''
  Phys.\ Lett.\ B {\bf 376}, 53 (1996).
  
\bibitem{Minwalla:1999px}
  S.~Minwalla, M.~Van Raamsdonk and N.~Seiberg,
  ``Noncommutative perturbative dynamics,''
  JHEP {\bf 0002}, 020 (2000)
  [arXiv:hep-th/9912072].

\bibitem{Hayakawa:1999zf}
  M.~Hayakawa,
  ``Perturbative analysis on infrared aspects of noncommutative QED on  R**4,''
  Phys.\ Lett.\ B {\bf 478}, 394 (2000)
  [arXiv:hep-th/9912094].


\bibitem{Gopakumar:2000zd}
  R.~Gopakumar, S.~Minwalla and A.~Strominger,
  ``Noncommutative solitons,''
  JHEP {\bf 0005}, 020 (2000)
  [arXiv:hep-th/0003160].
  
\bibitem{Zhou:2000xg}
  C.~G.~Zhou,
  ``Noncommutative scalar solitons at finite Theta,''
  arXiv:hep-th/0007255.

  
\bibitem{Hadasz:2001cn}
  L.~Hadasz, U.~Lindstrom, M.~Rocek and R.~von Unge,
  ``Noncommutative multisolitons: Moduli spaces, quantization, finite Theta
  effects and stability,''
  JHEP {\bf 0106}, 040 (2001)
  [arXiv:hep-th/0104017].
  
\bibitem{Durhuus:2001nj}
  B.~Durhuus, T.~Jonsson and R.~Nest,
  ``The existence and stability of noncommutative scalar solitons,''
  Commun.\ Math.\ Phys.\  {\bf 233}, 49 (2003)
  [arXiv:hep-th/0107121].





  
\bibitem{Nekrasov:1998ss}
  N.~Nekrasov and A.~S.~Schwarz,
  ``Instantons on noncommutative R**4 and (2,0) superconformal six  dimensional
  theory,''
  Commun.\ Math.\ Phys.\  {\bf 198}, 689 (1998)
  [arXiv:hep-th/9802068].
  
\bibitem{Furuuchi:1999kv}
  K.~Furuuchi,
  ``Instantons on noncommutative R**4 and projection operators,''
  Prog.\ Theor.\ Phys.\  {\bf 103}, 1043 (2000)
  [arXiv:hep-th/9912047].
  
\bibitem{Aganagic:2000mh}
  M.~Aganagic, R.~Gopakumar, S.~Minwalla and A.~Strominger,
  ``Unstable solitons in noncommutative gauge theory,''
  JHEP {\bf 0104}, 001 (2001)
  [arXiv:hep-th/0009142].
  
\bibitem{Furuuchi:2000dx}
  K.~Furuuchi,
  ``Dp-D(p+4) in noncommutative Yang-Mills,''
  JHEP {\bf 0103}, 033 (2001)
  [arXiv:hep-th/0010119].
  
\bibitem{Chu:2001cx}
  C.~S.~Chu, V.~V.~Khoze and G.~Travaglini,
  ``Notes on noncommutative instantons,''
  Nucl.\ Phys.\  B {\bf 621}, 101 (2002)
  [arXiv:hep-th/0108007].
  

  
  
\bibitem{Hamanaka:2001dr}
  M.~Hamanaka,
  ``ADHM/Nahm construction of localized solitons in noncommutative gauge
  theories,''
  Phys.\ Rev.\  D {\bf 65}, 085022 (2002)
  [arXiv:hep-th/0109070].
  
  
\bibitem{Gross:2000wc}
  D.~J.~Gross and N.~A.~Nekrasov,
  ``Monopoles and strings in noncommutative gauge theory,''
  JHEP {\bf 0007}, 034 (2000)
  [arXiv:hep-th/0005204].
  

  
\bibitem{Gross:2000ss}
  D.~J.~Gross and N.~A.~Nekrasov,
  ``Solitons in noncommutative gauge theory,''
  JHEP {\bf 0103}, 044 (2001)
  [arXiv:hep-th/0010090].








\bibitem{Polychronakos:2000zm}
  A.~P.~Polychronakos,
  ``Flux tube solutions in noncommutative gauge theories,''
  Phys.\ Lett.\  B {\bf 495}, 407 (2000)
  [arXiv:hep-th/0007043].
  
\bibitem{Jatkar:2000ei}
  D.~P.~Jatkar, G.~Mandal and S.~R.~Wadia,
  ``Nielsen-Olesen vortices in noncommutative Abelian Higgs model,''
  JHEP {\bf 0009}, 018 (2000)
  [arXiv:hep-th/0007078].
  

\bibitem{Gross:2000ph}
  D.~J.~Gross and N.~A.~Nekrasov,
  ``Dynamics of strings in noncommutative gauge theory,''
  JHEP {\bf 0010}, 021 (2000)
  [arXiv:hep-th/0007204].

  

  
\bibitem{Bak:2000ac}
  D.~Bak,
  ``Exact multi-vortex solutions in noncommutative Abelian-Higgs theory,''
  Phys.\ Lett.\  B {\bf 495}, 251 (2000)
  [arXiv:hep-th/0008204].
   

   

   

  
\bibitem{Lechtenfeld:2001aw}
  O.~Lechtenfeld and A.~D.~Popov,
  ``Noncommutative multi-solitons in 2+1 dimensions,''
  JHEP {\bf 0111}, 040 (2001)
  [arXiv:hep-th/0106213].
  

  
  
\bibitem{Chaichian:2004za}
  M.~Chaichian, P.~P.~Kulish, K.~Nishijima and A.~Tureanu,
  ``On a Lorentz-invariant interpretation of noncommutative space-time and  its
  implications on noncommutative QFT,''
  Phys.\ Lett.\  B {\bf 604}, 98 (2004)
  [arXiv:hep-th/0408069].

\bibitem{Wess:2003da}
  J.~Wess,
  ``Deformed coordinate spaces: Derivatives,''
  arXiv:hep-th/0408080.

\bibitem{Koch:2004ud}
  F.~Koch and E.~Tsouchnika,
  ``Construction of theta-Poincare algebras and their invariants on
  M(theta),''
  Nucl.\ Phys.\  B {\bf 717}, 387 (2005)
  [arXiv:hep-th/0409012].
  
\bibitem{Oeckl:2000eg}
  R.~Oeckl,
  ``Untwisting noncommutative R**d and the equivalence of quantum field
  theories,''
  Nucl.\ Phys.\  B {\bf 581}, 559 (2000)
  [arXiv:hep-th/0003018].

  
\bibitem{Chaichian:2004yh}
  M.~Chaichian, P.~Presnajder and A.~Tureanu,
  ``New concept of relativistic invariance in NC space-time: Twisted  Poincare
  symmetry and its implications,''
  Phys.\ Rev.\ Lett.\  {\bf 94}, 151602 (2005)
  [arXiv:hep-th/0409096].

\bibitem{Chaichian:2005yp}
  M.~Chaichian, K.~Nishijima and A.~Tureanu,
  ``An interpretation of noncommutative field theory in terms of a quantum
  shift,''
  Phys.\ Lett.\  B {\bf 633}, 129 (2006)
  [arXiv:hep-th/0511094].
  
\bibitem{Balachandran:2005eb}
  A.~P.~Balachandran, G.~Mangano, A.~Pinzul and S.~Vaidya,
  ``Spin and statistics on the Groenwald-Moyal plane: Pauli-forbidden  levels
  and transitions,''
  Int.\ J.\ Mod.\ Phys.\  A {\bf 21}, 3111 (2006)
  [arXiv:hep-th/0508002].

\bibitem{Balachandran:2005pn}
  A.~P.~Balachandran, A.~Pinzul and B.~A.~Qureshi,
  ``UV-IR mixing in non-commutative plane,''
  Phys.\ Lett.\  B {\bf 634}, 434 (2006)
  [arXiv:hep-th/0508151].
  
\bibitem{Lizzi:2006xi}
  F.~Lizzi, S.~Vaidya and P.~Vitale,
  ``Twisted conformal symmetry in noncommutative two-dimensional quantum field
  theory,''
  Phys.\ Rev.\  D {\bf 73}, 125020 (2006)
  [arXiv:hep-th/0601056].


\bibitem{Tureanu:2006pb}
  A.~Tureanu,
  ``Twist and spin-statistics relation in noncommutative quantum field
  theory,''
  Phys.\ Lett.\  B {\bf 638}, 296 (2006)
  [arXiv:hep-th/0603219].
  
\bibitem{Zahn:2006wt}
  J.~Zahn,
  ``Remarks on twisted noncommutative quantum field theory,''
  Phys.\ Rev.\  D {\bf 73}, 105005 (2006)
  [arXiv:hep-th/0603231].

  
\bibitem{Bu:2006ha}
  J.~G.~Bu, H.~C.~Kim, Y.~Lee, C.~H.~Vac and J.~H.~Yee,
  ``Noncommutative field theory from twisted Fock space,''
  Phys.\ Rev.\  D {\bf 73}, 125001 (2006)
  [arXiv:hep-th/0603251].

\bibitem{Abe:2006ig}
  Y.~Abe,
  ``Noncommutative quantization for noncommutative field theory,''
  Int.\ J.\ Mod.\ Phys.\  A {\bf 22}, 1181 (2007)
  [arXiv:hep-th/0606183].

\bibitem{Balachandran:2006pi}
  A.~P.~Balachandran, T.~R.~Govindarajan, G.~Mangano, A.~Pinzul, B.~A.~Qureshi and S.~Vaidya,
  ``Statistics and UV-IR mixing with twisted Poincare invariance,''
  Phys.\ Rev.\  D {\bf 75}, 045009 (2007)
  [arXiv:hep-th/0608179].

\bibitem{Fiore:2007vg}
  G.~Fiore and J.~Wess,
  ``On 'full' twisted Poincare' symmetry and QFT on Moyal-Weyl spaces,''
  Phys.\ Rev.\  D {\bf 75}, 105022 (2007)
  [arXiv:hep-th/0701078].

\bibitem{Joung:2007qv}
  E.~Joung and J.~Mourad,
  ``QFT with Twisted Poincar\'e Invariance and the Moyal Product,''
  JHEP {\bf 0705}, 098 (2007)
  [arXiv:hep-th/0703245].
  
\bibitem{Sasai:2007me}
  Y.~Sasai and N.~Sasakura,
  ``Braided quantum field theories and their symmetries,''
  Prog.\ Theor.\ Phys.\  {\bf 118}, 785 (2007)
  [arXiv:0704.0822 [hep-th]].
  
\bibitem{Riccardi:2007bj}
  M.~Riccardi and R.~J.~Szabo,
  ``Duality and Braiding in Twisted Quantum Field Theory,''
  arXiv:0711.1525 [hep-th].

  
\bibitem{Aschieri:2005yw}
  P.~Aschieri, C.~Blohmann, M.~Dimitrijevic, F.~Meyer, P.~Schupp and J.~Wess,
  ``A gravity theory on noncommutative spaces,''
  Class.\ Quant.\ Grav.\  {\bf 22}, 3511 (2005)
  [arXiv:hep-th/0504183].
  
\bibitem{Aschieri:2005zs}
  P.~Aschieri, M.~Dimitrijevic, F.~Meyer and J.~Wess,
  ``Noncommutative geometry and gravity,''
  Class.\ Quant.\ Grav.\  {\bf 23}, 1883 (2006)
  [arXiv:hep-th/0510059].  
  
\bibitem{Calmet:2005qm}
  X.~Calmet and A.~Kobakhidze,
  ``Noncommutative general relativity,''
  Phys.\ Rev.\  D {\bf 72}, 045010 (2005)
  [arXiv:hep-th/0506157].
  
\bibitem{Kobakhidze:2006kb}
  A.~Kobakhidze,
  ``Theta-twisted gravity,''
  [arXiv:hep-th/0603132]
  
\bibitem{Balachandran:2007kv}
  A.~P.~Balachandran, A.~Pinzul, B.~A.~Qureshi and S.~Vaidya,
  ``Twisted Gauge and Gravity Theories on the Groenewold-Moyal Plane,''
  arXiv:0708.0069 [hep-th].

\bibitem{Mukherjee:2006nd}
  P.~Mukherjee and A.~Saha,
  ``Comment on the first order noncommutative correction to gravity,''
  Phys.\ Rev.\  D {\bf 74}, 027702 (2006)
  [arXiv:hep-th/0605287].


\bibitem{Kurkcuoglu:2006iw}
  S.~Kurkcuoglu and C.~Saemann,
  ``Drinfeld twist and general relativity with fuzzy spaces,''
  Class.\ Quant.\ Grav.\  {\bf 24}, 291 (2007)
  [arXiv:hep-th/0606197].

\bibitem{Banerjee:2007th}
  R.~Banerjee, P.~Mukherjee and S.~Samanta,
  ``Lie algebraic Noncommutative Gravity,''
  Phys.\ Rev.\  D {\bf 75}, 125020 (2007)
  [arXiv:hep-th/0703128].

\bibitem{Oeckl:1999zu}
  R.~Oeckl,
  ``Braided quantum field theory,''
  Commun.\ Math.\ Phys.\  {\bf 217}, 451 (2001)
  [arXiv:hep-th/9906225].
  
\bibitem{Sasakura:2000vc}
  N.~Sasakura,
  ``Space-time uncertainty relation and Lorentz invariance,''
  JHEP {\bf 0005}, 015 (2000)
  [arXiv:hep-th/0001161].
  
\bibitem{Madore:2000en}
  J.~Madore, S.~Schraml, P.~Schupp and J.~Wess,
  ``Gauge theory on noncommutative spaces,''
  Eur.\ Phys.\ J.\  C {\bf 16}, 161 (2000)
  [arXiv:hep-th/0001203].
 
   
\bibitem{Imai:2000kq}
  S.~Imai and N.~Sasakura,
  ``Scalar field theories in a Lorentz-invariant three-dimensional
  noncommutative space-time,''
  JHEP {\bf 0009}, 032 (2000)
  [arXiv:hep-th/0005178].
      
\bibitem{Freidel:2005bb}
  L.~Freidel and E.~R.~Livine,
  ``Ponzano-Regge model revisited. III: Feynman diagrams and effective  field
  theory,''
  Class.\ Quant.\ Grav.\  {\bf 23}, 2021 (2006)
  [arXiv:hep-th/0502106].
   
\bibitem{Freidel:2005ec}
  L.~Freidel and S.~Majid,
  ``Noncommutative harmonic analysis, sampling theory and the Duflo map in  2+1
  quantum gravity,''
  arXiv:hep-th/0601004.

   
\bibitem{Derrick:1964ww}
  G.~H.~Derrick,
  ``Comments on nonlinear wave equations as models for elementary particles,''
  J.\ Math.\ Phys.\  {\bf 5}, 1252 (1964).

  
\end{thebibliography}
\end{document}